\documentclass[twocolumn,nofootinbib]{revtex4-1}
\pdfoutput=1
\usepackage{graphicx}
\usepackage{bm}
\usepackage{amssymb}
\usepackage{amsmath}
\usepackage{units}
\usepackage{array}
\usepackage{hyperref}
\usepackage{soul}
\usepackage{placeins}
\usepackage{float}
\usepackage{multirow}
\usepackage[T1]{fontenc}

\setcitestyle{square}

\def\fref{Fig.~\ref}
\def\sref{\S \ref}

\begin{document} 

\title{Dynamical Thin Disks}

%\correspondingauthor{John Ryan Westernacher-Schneider}

%\author[0000-0002-3047-7200]{John Ryan Westernacher-Schneider}
\author{John Ryan Westernacher-Schneider}
\email{john.westernacher.schneider@gmail.com}
\affiliation{Leiden Observatory, Leiden University, P.O. Box 9513, 2300 RA Leiden, The Netherlands}

\begin{abstract}
Thin disk accretion is often modeled in highly dynamical settings using the two-dimensional equations of viscous hydrodynamics, with viscosity representing unresolved turbulence. These equations are supposed to arise after vertical integration of the full three-dimensional equations of hydrodynamics, under the assumption of a geometrically thin disk with mirror symmetry about the midplane. However, in the dynamical context, vertical dynamics are neglected by incorrectly assuming instantaneous vertical hydrostatic equilibrium. The resulting errors in the local disk height couple to the horizontal dynamics through the $\alpha$-viscosity prescription and disk height-dependent gravitational softening models. Furthermore, the viscous terms in the horizontal equations are only complete if they are inserted after vertical integration, as if the system is actually two-dimensional. Since turbulence breaks mirror symmetry, it is more physically correct to insert a turbulence model at the three-dimensional level, and impose mirror symmetry only on average. Thus, some viscous terms are usually missing. With these considerations in mind, we revisit the vertical integration procedure, restricting to the regime of a Newtonian, non-self-gravitating disk. We obtain six evolution equations with only horizontal dependence, which determine the local vertical position and velocity of the disk surface, in addition to the usual surface densities of mass, horizontal momentum, and energy. This ``2.5-dimensional'' formulation opens the door to efficiently study vertical oscillations of thin disks in dynamical settings, and to improve the treatment of unresolved turbulence. As a demonstration, by including viscous stress at the three-dimensional level, we recover missing viscous terms which involve the vertical variables. We also propose a resummation of the vertically integrated gravitational force, which has a strikingly similar radial profile to a gravitational softening model advocated for in protoplanetary disk studies.
\end{abstract}

\maketitle
\section{Introduction}
% Uncomment to discover the column and text width in inches:
% \makeatletter
% \def\convertto#1#2{\strip@pt\dimexpr #2*65536/\number\dimexpr 1#1}
% \makeatother
% Text width: \convertto{in}{\the\textwidth}, Column width: \convertto{in}{\the\columnwidth}.

The treatment of geometrically thin disks \citep[][]{SS1973} historically embodied approximations about the vertical structure which go beyond thinness, including not-so-obvious ones such as the negligibility of vertical velocity and horizontal gradients of the disk scale height. This situation was pointed out early on \citep[e.g.][]{hoshi1977}, and the mathematical consistency of the approach was addressed clearly in~\cite{Abramowicz+1997}. This partly gave rise to the separate notion of a ``slim'' disk \citep[][]{abramowicz+1988}. The horizontal and vertical dynamics cannot be decoupled even for an eccentric steady disk \citep[][]{ogilvie2001}, let alone more dynamical scenarios like binary accretion.

For over 20 years, and continuing today, the original thin disk approach has been pushed beyond its regime of validity to study binary accretion \citep[e.g.][]{gunther+2002}. Due to its computational cost efficiency, and the large parameter space of binary accretion, two-dimensional simulations are unlikely to fall out of favour for the foreseeable future. Thus, improving the two-dimensional treatment is worthwhile. 

Turbulence is an essential aspect driving astrophysical accretion \citep[][]{SS1973}. In this work, we take the view that a turbulence model ought to be introduced at the three-dimensional level, since mirror symmetry about the disk midplane is only plausibly manifested in an averaged sense (i.e.~manifested at the level of a turbulence closure model). This is not the view always taken in the literature on thin disk turbulence, see e.g.~\cite{dubrulle1992}.

In this work, we seek to raise awareness of these issues. Restricting to the Newtonian, non-self-gravitating regime, we revisit the derivation of vertically integrated equations of motion assuming the disk is geometrical thin, mirror-symmetric about the midplane on average, has finite vertical extent, and allowing for multiple gravitating masses coplanar with the disk. The resulting equations are appropriate for shockwaves, large gradients, and other dynamical, nonaxisymmetric hydrodynamic phenomena characteristic of the binary accretion problem. We maintain only horizontal dependence of the variables, with the price being the evolution of two additional variables (the vertical position and velocity of the disk surface), as well as additional viscous terms involving the new variables. These equations should be useful for studying vertical oscillation phenomena, and to improve the treatment of unresolved turbulence. 

We use some inspiration from~\cite{Abramowicz+1997}, which provided a careful  mathematical appreciation of geometrically thin disks in the case where the vertical velocity is non-zero at leading order but the disk is steady. But we also would like to point out work treating thin disks in an asymptotic expansion in a characteristic aspect ratio \citep[e.g.][]{regev1983, kluzniak+2000, umurhan+2005}. This approach seems particularly instructive, and has been applied to study two-dimensional flows in axisymmetry \citep[e.g.][]{kluzniak+2000}, even including a treatment of magnetohydrodynamics (MHD) \citep[][]{cemeljic+2019}. Such an approach may also be fruitful for our current context of reducing non-axisymmetric problems to purely horizontal dependence, and may be useful for including, in a systematic way, additional physics such as radiative transfer \citep[e.g.][]{sadowski+2011, vorobyov+2017}, MHD, and associated advanced turbulence models \citep[e.g.][]{ogilvie2001, ogilvie2003, pessah+2006, miravet+2022}. We leave improvements of the cooling prescription, which might take steep horizontal gradients of the disk height into account, to future work.

Also note that axisymmetric versions of some of our results (e.g.~the disk height evolution equation) appear in~\cite{zhilkin+2021}; they consider a dynamical equation for the vertical velocity, but with a prescribed Gaussian vertical pressure profile and no turbulent viscosity terms. Nontrivial vertical dynamics were also considered in specialized coordinates to study eccentric disk evolution in~\cite{ogilvie2001}. In work on eccentric tidal disruption event disks, vertical dynamics has also been considered \citep[see e.g.][]{lynch+2021}. In this work, we leave the horizontal coordinates arbitrary.

Vertically integrated gravity models often have a free parameter (the ``softening length'') which regulates the divergence at point masses and represents finite disk thickness. To eliminate the softening length as a free parameter in simulations, we propose a resummation of vertically integrated gravity which depends on the local disk height $H$ (measured from the midplane). For small aspect ratios, this proposal bears a striking resemblance to Plummer models with softening length $\sim 0.6 H$, which is a model advocated for in protoplanetary studies based on comparisons to three-dimensional calculations \citep[e.g.][]{Mueller+2012}.

In section \sref{sec:prelim}, we provide mathematical preliminaries and the final ``2.5-dimensional'' equations; we derive them in subsequent sections.

%
%%
%%%
%%
%

\section{preliminaries and final equations} \label{sec:prelim}
Consider a disk of gas situated in the plane $z=0$ in a coordinate system $\lbrace u,w,z \rbrace$, where $\lbrace u, w \rbrace$ represent arbitrary horizontal coordinates independent of $z$, e.g.~Cartesian $\lbrace x,y \rbrace$ or cylindrical $\lbrace r, \theta \rbrace$, and $z$ is the Cartesian vertical coordinate. In everything that follows, $r$ will denote the cylindrical radial coordinate. In index notation, we denote spatial directions with indices $\lbrace i,j,k,...\rbrace$ and the horizontal directions with capital indices $\lbrace I,J,K,...\rbrace$. In terms of metric language, our coordinate assumptions mean a flat space with line element
\begin{eqnarray}
    ds^2 = g_{IJ}dx^I dx^J + dz^2 \label{eq:metric}
\end{eqnarray}
where the horizontal metric coefficients $g_{IJ}$ are independent of $z$. The origin is the center of mass of a system of point masses whose orbits also lie in the plane $z=0$. We assume the disk has reflection symmetry about the $z=0$ plane in an averaged sense by inserting a viscous turbulence model at the 3-dimensional level, with all the fluid variables representing appropriately defined mean values \citep[e.g.~Favre averages,][]{favre1969}. Reflection symmetry implies that scalar quantities and horizontal velocities are even functions about $z=0$, whereas the vertical velocity $v^z$ is an odd function about $z=0$.

At a given $(t,u,w)$, where $t$ is time, the disk extends vertically to $z=\pm H(t,u,w)$, where $H$ is the local disk height. $H/r$ is the local aspect ratio. The disk is thin in the sense that $z/r\ll1$ inside it, so $z/r$ is a sensible small parameter to facilitate perturbative solutions. The surfaces at $z=\pm H(t,u,w)$ are treated as fluid-vacuum interfaces. We thus weight the pressure and density by a distribution $W$ built from step functions $\Theta$,
\begin{eqnarray}
W = \Theta\left(z + H(t,u,w)\right) - \Theta\left(z - H(t,u,w)\right).
\end{eqnarray}
$W$ is zero in vacuum and unity inside the disk. Its purpose is to truncate vertical integrals in a computationally elegant way. Note that under conditions of vertical hydrostatic balance, a polytropic equation of state in $z$ yields fluid-vacuum interfaces, whereas the specific case of a $\Gamma=1$ polytrope has infinite vertical extent. Morally speaking, one could view the vacuum boundary as an approximate notion, with $H$ actually being a scale height. With mirror symmetry, our ansatz for the fluid variables are therefore given by
\begin{eqnarray}
\rho &=& \left[\rho_0 + \rho_2 (z/r)^2 + \mathcal{O}(z/r)^4\right] W \\
P &=& \left[P_0 + P_2 (z/r)^2 + \mathcal{O}(z/r)^4\right] W \\
v^I &=& v^I_0 + v^I_2 (z/r)^2 + \mathcal{O}(z/r)^4 \\
v^z &=& v^z_1(z/r) + \mathcal{O}(z/r)^3,
\end{eqnarray}
where the integer subscripts denote the coefficients of the respective terms in the power series in $z/r$. Such coefficients are dependent upon time and horizontal position only, whereas $W$ depends on horizontal position and time (via dependence on $H$) as well as on the vertical coordinate.

We present the final equations now, and derive them in subsequent sections. For cleanliness we separate the viscous terms, which should be understood as appearing on the right-hand side of their respective equations of motion. ${v_{}}_H\equiv v_z\vert_H$ is the vertical velocity evaluated at $z=H$. The zeroth-order vertically integrated mass density, pressure, and energy density are $\Sigma \equiv \rho_0 H$, $\mathcal{P} \equiv P_0 H$, $\mathcal{E} \equiv \Sigma \epsilon_0 + (1/2)\Sigma v_0^2$. Subscripts $n$ index the gravitating point masses, all situated in the midplane $z=0$. $D_n$ is the in-plane distance from a field point to the $n$th point mass in the system, and $\hat{D}_n$ is the corresponding outward-pointing horizontal unit vector. $\tilde{\Omega}^2$ is a multi-body generalization of the squared Keplerian frequency, equal to the sum of squared Keplerian frequencies for individual bodies. $^{\rm 2D}\tau_{IJ} \equiv \Sigma\nu(\nabla_I v_{0,J} + \nabla_J v_{0,I} - (2/3) g_{IJ} \nabla_K v_0^K) + \Sigma \lambda \nabla_K v_0^K$ is the purely two-dimensional viscous stress tensor, with $\nu$ and $\lambda$ the kinematic shear and bulk viscosities. Unless a polytropic equation of state is imposed, the energy equation must be supplemented by a cooling prescription, which we leave to future work.
\begin{widetext}
\begin{equation*}
\hspace{-1cm}
\begin{aligned}[c]
\phantom{\Biggl(}\text{Mass}&\text{:} \\
\phantom{\Biggl(}\text{Horizontal momentum}&\text{:} \\
\phantom{\Biggl(}\text{Energy}&\text{:} \\
\phantom{\Biggl(}\text{Vertical momentum}&\text{:} \\
\phantom{\Biggl(}\text{Disk height}&\text{:} \\
\phantom{\Biggl(}\text{Resummation of gravity}&\text{:} \\
\phantom{\Biggl(}\text{Horizontal viscosity}&\text{:} \\
\phantom{\Biggl(}\text{Energy viscosity}&\text{:} \\
\phantom{\Biggl(} \\
\phantom{\Biggl(}\text{Vertical viscosity}&\text{:} \\
\phantom{\Biggl(}
\end{aligned}
\hspace{1cm}
\begin{aligned}[c]
\phantom{\Biggl(}\partial_t \Sigma + \nabla_J \left( \Sigma v_0^J \right) &= 0 \\
\phantom{\Biggl(} \partial_t \left( \Sigma v_{0,I} \right) + \nabla_J \left( \Sigma v_0^J v_{0,I} \right) &= -\partial_I \mathcal{P} - \Sigma \sum_n \frac{GM_n}{D_n^2} \hat{D}_n \\
\phantom{\Biggl(} \partial_t \left( \mathcal{E} \right) + \nabla_J \left[ \left( \mathcal{E} + \mathcal{P} \right) v_0^J \right]
&= -\Sigma \sum_n \frac{GM}{D_n^2} \hat{D}_n \cdot \vec{v}_0 \\
\phantom{\Biggl(} \partial_t \left( \Sigma {v_{}}_H\right) + \nabla_J \left( \Sigma v_0^J {v_{}}_H \right) &= \frac{2\mathcal{P}}{H} - \Sigma \tilde{\Omega}^2 H \\
\phantom{\Biggl(} \frac{dH}{dt} = \partial_t H + v_0^J \partial_J H &= {v_{}}_H \\
\phantom{\Biggl(} \Sigma \sum_n \frac{GM}{D_n^2} &\longrightarrow \Sigma \sum_n \frac{GM}{D_n^2} \frac{1}{\sqrt{1+(H/D_n)^2}}\\
\phantom{\Biggl(} \int_0^\infty\!\! dz \nabla_j \tau^j_I &\simeq \nabla_J \left(^{\rm 2D}\tau^J_I\right) - \nabla_I \left[ \frac{\Sigma}{H} \left(\frac{2}{3}\nu - \lambda\right) {v_{}}_H \right] \\
\phantom{\Biggl(} \int_0^\infty dz \nabla_j \left(v^i\tau^j_i\right)
&\simeq \nabla_J \left(v^I\! \left(^{\rm 2D}\tau^J_I\right)\right)
- \nabla_J \left(\frac{\Sigma}{H} \left(\frac{2}{3}\nu - \lambda\right) v_0^J {v_{}}_H \right) \nonumber\\
\phantom{\Biggl(}&+ \nabla_J \left[ \frac{\Sigma}{H} \nu \frac{{v_{}}_H}{3} \left[ H \nabla^J {v_{}}_H \!-\! {v_{}}_H \nabla^J H \right]\! \right]\\
\phantom{\Biggl(} \int_0^\infty dz\nabla_j \tau^j_z &\simeq
\nabla_J \left[ \frac{1}{2} \frac{\Sigma}{H} \nu \left( H \nabla^J {v_{}}_H - {v_{}}_H \nabla^J H \right)\right] \nonumber\\
\phantom{\Biggl(}&- \frac{\Sigma}{H} \left\lbrace \left(\frac{4}{3}\nu + \lambda \right) \frac{{v_{}}_H}{H} - \left(\frac{2}{3}\nu - \lambda\right) \nabla_J v_0^J \right\rbrace
\end{aligned}
\end{equation*}
\end{widetext}
$\phantom{o}$\\
$\phantom{o}$\\
$\phantom{o}$\\

\section{Vertical dynamics}

In order to solve for the vertical structure, we must solve the equation of momentum conservation along $z$,
\begin{eqnarray}
\partial_t \left( \rho v_z \right) + \nabla_j \left( \rho v^j v_z \right) = \rho g_z -\partial_z P, \label{eq:zmomcons}
\end{eqnarray}
where $g_z$ is the vertical gravitational acceleration $-\partial_z \Phi$. For cleanliness, we consider the viscous terms later.

For the gravity term, we sum the potentials of $N$ point masses situated in the plane,
\begin{eqnarray}
\Phi = -\sum_{n=1}^N \frac{G M_n}{R_n}, \label{eq:potl}
\end{eqnarray}
where $G$ is Newton's constant, $M_n$ is the $n$th mass, and $R_n$ is the distance between the field point $(u,w,z)$ and the mass location $(u_n, w_n, 0)$, given as
\begin{eqnarray}
R_n = \sqrt{z^2 + D_n^2}, \label{eq:nsep}
\end{eqnarray}
where $D_n$ is the distance between the points $(u,w,0)$ and $(u_n,w_n,0)$.
The gravitational acceleration along $z$ is 
\begin{eqnarray}
g_z &=& - \partial_z \Phi \nonumber\\
&=& - \sum_{n=1}^N \frac{G M_n}{R_n^2} \partial_z R_n \nonumber\\
&=& - \sum_{n=1}^N \frac{G M_n}{D_n^2} \frac{z}{D_n} + \mathcal{O}\left( \left(\frac{z}{D_n}\right)^2 \frac{z}{r} \right) \\
&\simeq& -\tilde{\Omega}^2 z. \label{eq:gz}
\end{eqnarray}
We defined $\tilde{\Omega}^2 \equiv \sum_{n=1}^N (GM_n/D_n^3)$ as a multi-body generalization of the squared Keplerian frequency. Plugging everything into Eq.~\eqref{eq:zmomcons} and keeping lowest order terms yields
\begin{eqnarray}
&\partial_t& \left( \rho_0 W v_{1,z}\frac{z}{r}\right)
+
\nabla_J \left( \rho_0 W v_0^J v_{1,z} \frac{z}{r} \right) \nonumber\\
&+&
\partial_z \left( \rho_0 W v_{1,z}^2 (z/r)^2 \right)
= 
-\rho_0 W \tilde{\Omega}^2 z
-
\partial_z\left( P_0 W \right)\!. \phantom{oooo}
\label{eq:vertstruct}
\end{eqnarray}
Next, integrate Eq.~\eqref{eq:vertstruct} vertically over $z\in[0,\infty)$. The $\partial_z$ term on the left-hand side vanishes, and we obtain
\begin{eqnarray}
\partial_t \left( \Sigma v_z\vert_H\right)
+
\nabla_J \left( \Sigma v_0^J v_z\vert_H \right)
=
\frac{2\mathcal{P}}{H}
-
\Sigma \tilde{\Omega}^2 H
\label{eq:vertvert}
\end{eqnarray}
where $\Sigma \equiv H\rho_0$ and $\mathcal{P} \equiv HP_0$ and $v_z\vert_H$ is the vertical velocity evaluated at the disk surface $z=H$. Note $\Sigma v_z\vert_H$ is twice the vertically integrated vertical momentum.

When $v_z\vert_H =0$, hydrostatic balance is achieved:
\begin{eqnarray}
H = \sqrt{\frac{2\mathcal{P}}{\Sigma}} \frac{1}{\tilde{\Omega}}.
\end{eqnarray}
But if $H$ depends on time $t$, then $v_z\vert_H$ cannot be zero. Roughly speaking, we expect by definition that
\begin{eqnarray}
\partial_t H \sim v_z\vert_{H}.
\end{eqnarray}
Now suppose shocks are generated in the disk, and that they tend to be smeared out by turbulent eddies (of size $\sim H$) to a width $\sim H$. Since thin disks are highly supersonic, those spatial variations would propagate horizontally at roughly Keplerian speed $v_K$ (i.e.~at speed $\sim v_K + c_s \sim v_K + v_K/\mathcal{M} \sim v_K$ where $\mathcal{M}=v_K/c_s$ is the Mach number and $c_s$ is the sound speed). Thus, $\mathcal{O}(1)$-changes in $H$ occur on a time scale $\sim H/v_K$, implying
\begin{eqnarray}
\partial_t H \sim \frac{H}{H/v_K} = v_K \sim v_z\vert_H. \label{eq:cantneglectvz}
\end{eqnarray}
Thus, the terms in Eq.~\eqref{eq:vertstruct} involving $v_z\vert_H$ are dominant when steep gradients are propagating.\footnote{Note $\partial_t \sim v_K \partial_J$} Thus, in intermediate conditions, all terms can be of the same order. It is therefore generally unjustified in a dynamical setting to ignore the terms involving the vertical velocity. 

The integrated vertical momentum equation is an evolution equation for $v_z\vert_H$, but we still have a 2-dimensional dependence of the fluid variables. We must also relate the fluid variables to $H$ when $v_z\vert_H\neq 0$. To determine $H$, we can employ the free surface boundary conditions, which include that the surface moves according to the fluid velocity at $z=H$:
\begin{eqnarray}
\frac{dH}{dt} = \partial_t H + v_0^J \partial_J H = v_z\vert_H. \label{eq:Hevo}
\end{eqnarray}
Using the outward-pointing normal to the surface, $\hat{N}$, this equation of motion~\eqref{eq:Hevo} can be derived by considering a flat, infinitesimal segment of surface, with inclination with respect to the vertical direction determined by $\hat{N}\cdot\hat{z}$, moving due to the full fluid velocity $\vec{v}\vert_H$ in the direction determined by $\hat{N}\cdot\vec{v}\vert_H$ by an amount $\hat{N}\cdot\vec{v}\vert_H dt$ in time $dt$. Then at fixed horizontal coordinates, the surface moves vertically by an amount
\begin{eqnarray}
dH \hat{N}\cdot\hat{z} = \hat{N}\cdot\vec{v}\vert_H dt. \label{eq:Hevoprep}
\end{eqnarray}
The outward-pointing normal is given in Cartesian coordinates as
\begin{eqnarray}
\hat{N} = \frac{[-\partial_x H, -\partial_y H, 1]^T}{\sqrt{1+ (\partial_x H)^2 + (\partial_y H)^2}}.
\end{eqnarray}
Using this form, Eq.~\eqref{eq:Hevo} follows from Eq.~\eqref{eq:Hevoprep}. Note that Eq.~\eqref{eq:Hevo} is a non-axisymmetric generalization of a formula written down in~\cite{zhilkin+2021} (see their equation 20). Those authors also consider the dynamical equation for the vertical velocity (their equation 17), but with a prescribed Gaussian vertical pressure profile (and thus no fluid-vacuum interface) and no turbulent viscosity terms.

In a two-dimensional calculation, one could evolve $v_z\vert_H(t,u,w)$ and $H(t,u,w)$ using Eqs.~\eqref{eq:vertstruct} \&~\eqref{eq:Hevo}, respectively, in addition to the remaining fluid equations. Standard shock-capturing methods can be applied readily to Eq.~\eqref{eq:vertstruct}, whereas we anticipate that Eq.~\eqref{eq:Hevo} just needs a slope-limited differencing operator applied to $H$.

%
%%
%%%
%%
%
\subsection{Viscous Terms} \label{sec:vertvisc}
The viscous term in the vertical momentum equation would appear on the right-hand side of Eq.~\eqref{eq:zmomcons} as $\nabla_j \tau^j_z$, where 
\begin{eqnarray}
\!\!\!\!\!\!\!\tau_{ij} = \rho\! \left[ \nu\! \left( \nabla_i v_j\! +\! \nabla_j v_i\! -\! \frac{2}{3} g_{ij} \nabla_k v^k \right)\! +\! \lambda g_{ij} \nabla_k v^k \right]\!\!, \label{eq:stress}
\end{eqnarray}
and where we defined the kinematic viscosity $\nu$ and the bulk viscosity per density $ \lambda$ (which we will simply call the bulk viscosity). The bulk viscosity $\lambda$ controls the trace of the viscous stress tensor $\tau_{ij}$. It has been suggested that bulk viscosity can represent radiative damping \citep[][]{papaloizou+1977} or a stabilizing influence similar to the finite relaxation time scale of unresolved MHD turbulence \citep[][]{ogilvie2001}.

In the vertical structure equation, on the right-hand side we have
\begin{eqnarray}
&\nabla_j& \tau^j_z = \nabla_J \tau^J_z + \partial_z \tau^z_z \nonumber\\
&=& \nabla_J \left[ \rho \nu \left( \nabla^J v_z + \partial_z v^J \right)\right] \nonumber\\
&+&
\partial_z \left[ \rho \left[ \left( \frac{4}{3}\nu + \lambda \right) \partial_z v^z - \left( \frac{2}{3}\nu - \lambda \right) \nabla_J v^J \right] \right]\!, \phantom{ooo}
\end{eqnarray}
where we used $\nabla_z \tau^z_z = \partial_z \tau^z_z$ and $\nabla_z v^z = \partial_z v^z$ (since the Christoffel symbols obey $\Gamma^z_{zj}=0=\Gamma^j_{zz}$). Integrating over $z\in[0,\infty)$ yields
\begin{eqnarray}
&\int_0^\infty& dz\nabla_j \tau^j_z \simeq \nonumber\\
&\nabla_J& \left[ \frac{1}{2} \frac{\Sigma}{H} \nu \left( H \nabla^J (v_z\vert_H) - v_z\vert_H \nabla^J H + 2 v_2^J \left(\frac{H}{r}\right)^{\!\! 2} \right)\right] \nonumber\\
&-& \frac{\Sigma}{H} \left\lbrace \left(\frac{4}{3}\nu + \lambda \right) \frac{v_z\vert_H}{H} - \left(\frac{2}{3}\nu - \lambda\right) \nabla_J v_0^J \right\rbrace \label{eq:vertvisc}
\end{eqnarray}
Owing to two derivatives, a higher order coefficient in the power series for velocity now appears, i.e.~$ v_2^J $. As long as $v_2^J$ is at most the same order as $v_0^J$, the term involving $v_2^J$ should be higher order. For example, in relaxed conditions, where horizontal derivatives $\nabla\sim 1/r$, Eq.~\eqref{eq:Hevo} implies that $v_z\vert_H \sim v_0^J (H/r)$. Thus, all terms in the square brackets in Eq.~\eqref{eq:vertvisc} are of the same order, and smaller than the terms in curly brackets by $(H/r)^2$. In dynamic conditions, where $\nabla\sim 1/H$ and $v_z\vert_H \sim v_0^J$, the term involving $v_2^J$ is smaller than all other terms in Eq.~\eqref{eq:vertvisc} by $(H/r)^2$. Thus, so long as $v_2^J$ is at most the order of $v_0^J$, we can neglect the $v_2^J$ term. This condition will certainly not hold in general, since axisymmetric calculations have shown that horizontal velocities can flip direction between $z=0$ and $z=H$, even resulting in coexisting midplane backflow and surface inflow \citep[][]{urpin1984, kley+1992, kluzniak+2000, regev+2002, mishra+2022}. Any hope of accounting for this sort of behavior in a vertically integrated context requires carrying out expansions to sufficiently high order that the velocity field can be reconstructed with enough vertical detail. Such an approach would still be two-dimensional, but with additional expansion coefficients being evolved.

%
%%
%%%
%%%%
%%%
%%
%

%
%%
%%%
%%%%
%%%
%%
%
\section{Horizontal dynamics}

\subsection{Local conservation of mass}

At lowest order, the continuity equation reads
\begin{eqnarray}
\partial_t \left( \rho_0 W \right) + \nabla_J \left( \rho_0 v_0^J \right) = - \partial_z \left( \rho_0 W v^z \right), \label{eq:mass3d}
\end{eqnarray}
where a source term comes from the 3-dimensional divergence, encoding the vertical dynamics. Integrating over $z\in[0,\infty)$ yields
\begin{eqnarray}
\partial_t \Sigma + \nabla_J \left( \Sigma v_0^J \right) = 0. \label{eq:mass2d}
\end{eqnarray}

\subsection{Local conservation of momentum}

At lowest order, the horizontal momentum equation reads
\begin{eqnarray}
\partial_t \left( \rho_0 W v_{0,I} \right) &+& \nabla_J \left( \rho_0 W v_0^J v_{0,I} \right) + \partial_z \left( \rho_0 W v^z v_{0,I} \right) \nonumber\\
&=& -\partial_I \left( P_0 W \right) + \rho_0 W g_I. \label{eq:mom3d}
\end{eqnarray}
Integrating over $z\in [0,\infty)$ yields
\begin{eqnarray}
\partial_t \left( \Sigma v_{0,I} \right) + \nabla_J \left( \Sigma v_0^J v_{0,I} \right) &=& -\partial_I \mathcal{P} \nonumber\\
&-& \Sigma \sum_n \frac{GM_n}{D_n^2} \hat{D}_n, \label{eq:mom2d}
\end{eqnarray}
where $\hat{D}_n$ is a unit vector pointing away from the $n$th point mass. At lowest order, the gravitational force has purely Newtonian form. Often the gravitational force is instead approximated as that arising from a Plummer potential, which one could argue is a model of the gravitational force at beyond leading order in $z/r$ \citep[see e.g.][]{hure+2009, Mueller+2012}.

\subsection{Proposed resummation of the gravitational force}

\begin{figure}
\centering
\includegraphics[width=0.47\textwidth]{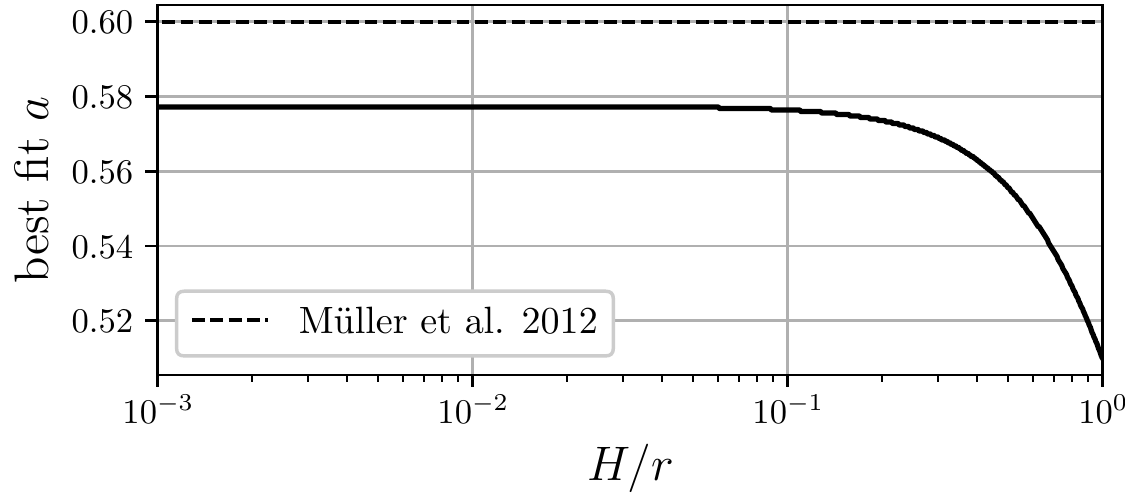}
\caption{Best fits of Eq.~\eqref{eq:gresum} with a Plummer model with softening length $aH$ for different aspect ratios.} \label{fig:fit_plummer}
\end{figure}

Set $\rho = \rho_0 W$ and leave the gravitational acceleration exact, as $\sum_n GM_n D_n/R_n^3$, then integrate. This yields
\begin{eqnarray}
\rho_0 \sum_n \frac{GM}{D_n^2} &\times&\int_0^H dz (1 + (z/D_n)^2)^{-3/2} \nonumber\\
&=& \Sigma \sum_n \frac{GM}{D_n^2} \frac{1}{\sqrt{1+(H/D_n)^2}}. \label{eq:gresum}
\end{eqnarray}
Compared to a Plummer model with softening length $aH$, this proposed force law is best fit by a Plummer model at small aspect ratios when $a\simeq 0.58$, which is a similar softening length that has been advocated for use when modeling non-self-gravitating disks in planetary studies \citep[e.g.~$a\simeq0.6$, see][and references therein]{Mueller+2012}. The best-fitting value of $a$ decreases toward $0.51$ as the aspect ratio goes to 1 and the thin disk approach breaks down -- see \fref{fig:fit_plummer}. 

Regulating Eq.~\eqref{eq:gresum} at $D_n=0$ is another matter. One can, for example, transition smoothly to a traditional Plummer model with fixed softening length at very small radii, such as within numerical sinks which represent accreting bodies.

\subsection{Local conservation of energy}

The energy equation reads
\begin{eqnarray}
\partial_t\! \left(\!\rho \epsilon + \frac{1}{2}\rho v^2\! \right)\! +\! \nabla_j\! \left[\! \left( \rho\epsilon\! +\! \frac{1}{2}\rho v^2\! +\! P \right)\! v^j \right]\! =\! \rho v^j g_j.\phantom{oo} \label{eq:enfull}
\end{eqnarray}
At lowest order, we obtain
\begin{eqnarray}
&\partial_t& \left( \rho_0 W \epsilon_0 + \frac{1}{2} \rho_0 W v_0^2 \right) \nonumber\\
&+& \nabla_J \left[ W \left( \rho_0 \epsilon_0
+ \frac{1}{2} \rho_0 v_0^2  + P_0\right) v_0^J \right] \nonumber\\
&+& \partial_z \left[ W \left( \rho_0 \epsilon_0
+ \frac{1}{2} \rho_0 v_0^2  + P_0\right) v^z \right]
= \rho_0 W v_0^J g_J.\phantom{ooo} \label{eq:en3d}
\end{eqnarray}
Integration over $z\in[0,\infty)$ yields
\begin{eqnarray}
\partial_t \left( \Sigma \epsilon_0 + \frac{1}{2} \Sigma v_0^2 \right) &+& \nabla_J \left[ \left( \Sigma \epsilon_0 + \frac{1}{2} \Sigma v_0^2 + \mathcal{P} \right) v_0^J \right] \nonumber\\
&=& -\Sigma \sum_n \frac{GM}{D_n^2} \hat{D}_n \cdot \vec{v}_0. \label{eq:en2d}
\end{eqnarray}
The gravitational acceleration term can also use the resummed form given by Eq~\eqref{eq:gresum}. A radiative cooling prescription should be inserted on the right-hand side. A careful computation of radiative cooling, appropriate for dynamic disk conditions, is left to future work.

%
%%
%%%
%%%%
%%%
%%
%
\subsection{Viscosity}
Now we compute the vertical integration of the viscous terms. The derivations are tedious but straightforward, so we omit them. We define the purely horizontal vertically integrated viscous stress as
\begin{eqnarray}
^{\rm 2D}\!\tau^J_I\! \equiv\! \Sigma \nu\! \left(\!\nabla^J\! v_{0,I}\! +\! \nabla_{\!I} v^J_0 \!-\! \frac{2}{3} \delta^J_I \nabla_K v_0^K\!\right)\! +\! \Sigma\lambda \nabla_K v_0^K\!. \phantom{ooo}
\end{eqnarray}
The viscous term on the right-hand side of the horizontal momentum equation is:
\begin{eqnarray}
\int_0^\infty\!\! dz \nabla_j \tau^j_I \!\simeq\! \nabla_J \left(^{\rm 2D}\tau^J_I\right) \!-\! \nabla_I \left[ \frac{\Sigma}{H} \left(\frac{2}{3}\nu - \lambda\right) v^z\vert_H \right]\!. \phantom{ooo}
\end{eqnarray}

For the energy equation, the viscous term on the right-hand side is:
\begin{eqnarray}
&&\int_0^\infty dz \nabla_j \left(v^i\tau^j_i\right) \nonumber\\
&\simeq& \nabla_J \left(v^I\! \left(^{\rm 2D}\tau^J_I\right)\right)
- \nabla_J \left(\frac{\Sigma}{H} \left(\frac{2}{3}\nu - \lambda\right) v_0^J v_z\vert_H \right) \nonumber\\
&+& \nabla_J \left[ \frac{\Sigma}{H} \nu \frac{v_z}{3}\vert_H \left[ H \nabla^J (v_z\vert_H) \!-\! v_z\vert_H \nabla^J H \!+\! 2 v_2^J \left( \!\frac{H}{r}\! \right)^{\!\! 2}\right]\! \right]\!.\nonumber\\
\end{eqnarray}
Using the same arguments as in \sref{sec:vertvisc}, we neglect the $v_2^J$ term which closes the system.

%
%%
%%%
%%%%
%%%
%%
%

\section{Conclusions \& Outlook} \label{sec:conclude}
The thin disk equations traditionally used in highly dynamical calculations, such as binary accretion, neglect vertical dynamics. Turbulence models are also typically inserted after vertical integration, which does not respect the fact that turbulence breaks mirror symmetry at the three-dimensional level. Meanwhile, the two-dimensional approach is computationally indispensable, since it allows for efficient exploration of the large parameter space of thin disk accretion problems. 

In this work, we sought to raise awareness of these issues. We revisited the reduction of a thin disk to two dimensions, and obtained hydrodynamic equations of motion suited for thin disks with shockwaves and other dynamic phenomena. The resulting system is still effectively two-dimensional in the variable dependence, and the local vertical position and velocity of the disk surface are new field variables. We demonstrated that insertion of a turbulence model at the three-dimensional level results in new viscous terms after vertical integration. We also proposed a resummation of the horizontal gravitational force, which acts to soften the force law (which is physically required at next-to-leading order in the disk aspect ratio) in a way which eliminates the softening length as a free parameter. A Plummer model with softening length $\sim0.6H$, which has been advocated for in protoplanetary studies, bears a striking resemblance to our proposed resummation of the gravitational force for small aspect ratios, and differs increasingly as the aspect ratio approaches unity. An improvement of radiative cooling, appropriate for e.g.~large gradients of the disk thickness, is left to future work.

We plan to implement our new ``2.5''-dimensional system of equations in binary accretion simulations, and check whether known results are sensitive to the neglect of vertical dynamics and a proper three-dimensional treatment of turbulence. Vertical oscillation phenomena in such systems, and their observational signatures, will also be explored. In the future, detailed vertical structure and nontrivial meridional flow can likely be resolved with effectively two-dimensional simulations if the equations of motion are solved to sufficiently high order in the disk aspect ratio, the expense of which is increased equation complexity and the evolution of additional variables (the higher-order coefficients in the $(z/r)$-expansions of the fluid variables).

\renewcommand\bibname{References}

\bibliographystyle{unsrt}
\bibliography{cbd}

\begin{thebibliography}{10}

\bibitem{SS1973}
N.~I. {Shakura} and R.~A. {Sunyaev}.
\newblock {Reprint of 1973A\&A....24..337S. Black holes in binary systems.
  Observational appearance.}
\newblock {\em A\&A}, 500:33--51, June 1973.

\bibitem{hoshi1977}
R.~{H{\={o}}shi}.
\newblock {Basic Properties of a Stationary Accretion Disk Surrounding a Black
  Hole}.
\newblock {\em Progress of Theoretical Physics}, 58(4):1191--1204, October
  1977.

\bibitem{Abramowicz+1997}
M.~A. {Abramowicz}, A.~{Lanza}, and M.~J. {Percival}.
\newblock {Accretion Disks around Kerr Black Holes: Vertical Equilibrium
  Revisited}.
\newblock {\em \apj}, 479(1):179--183, April 1997.

\bibitem{abramowicz+1988}
M.~A. {Abramowicz}, B.~{Czerny}, J.~P. {Lasota}, and E.~{Szuszkiewicz}.
\newblock {Slim Accretion Disks}.
\newblock {\em \apj}, 332:646, September 1988.

\bibitem{ogilvie2001}
G.~I. {Ogilvie}.
\newblock {Non-linear fluid dynamics of eccentric discs}.
\newblock {\em MNRAS}, 325(1):231--248, July 2001.

\bibitem{gunther+2002}
R.~{G{\"u}nther} and W.~{Kley}.
\newblock {Circumbinary disk evolution}.
\newblock {\em A\&A}, 387:550--559, May 2002.

\bibitem{dubrulle1992}
B.~{Dubrulle}.
\newblock {A turbulent closure model for thin accretion disks}.
\newblock {\em A\&A}, 266(1):592--604, December 1992.

\bibitem{regev1983}
O.~{Regev}.
\newblock {The disk-star boundary layer and its effect on the accretion disk
  structure}.
\newblock {\em A\&A}, 126(1):146--151, September 1983.

\bibitem{kluzniak+2000}
W.~{Klu{\'z}niak} and D.~{Kita}.
\newblock {Three-dimensional structure of an alpha accretion disk}.
\newblock {\em arXiv e-prints}, pages astro--ph/0006266, June 2000.

\bibitem{umurhan+2005}
O.~M. {Umurhan} and G.~{Shaviv}.
\newblock {On the nature of the hydrodynamic stability of accretion disks}.
\newblock {\em A\&A}, 432(2):L31--L34, March 2005.

\bibitem{cemeljic+2019}
M.~{\v{C}}emelji{\'c}, W.~{Klu{\'z}niak}, and V.~{Parthasarathy}.
\newblock {A magnetized thin accretion disk: numerical simulations compared
  with asymptotic expansion}.
\newblock {\em arXiv e-prints}, page arXiv:1907.12592, July 2019.

\bibitem{sadowski+2011}
A.~{S{\k{a}}dowski}, M.~{Abramowicz}, M.~{Bursa}, W.~{Klu{\'z}niak}, J.~P.
  {Lasota}, and A.~{R{\'o}{\.z}a{\'n}ska}.
\newblock {Relativistic slim disks with vertical structure}.
\newblock {\em A\&A}, 527:A17, March 2011.

\bibitem{vorobyov+2017}
Eduard~I. {Vorobyov} and Yaroslav~N. {Pavlyuchenkov}.
\newblock {Improving the thin-disk models of circumstellar disk evolution. The
  2+1-dimensional model}.
\newblock {\em A\&A}, 606:A5, September 2017.

\bibitem{ogilvie2003}
G.~I. {Ogilvie}.
\newblock {On the dynamics of magnetorotational turbulent stresses}.
\newblock {\em MNRAS}, 340(3):969--982, April 2003.

\bibitem{pessah+2006}
Martin~E. {Pessah}, Chi-Kwan {Chan}, and Dimitrios {Psaltis}.
\newblock {Local Model for Angular-Momentum Transport in Accretion Disks Driven
  by the Magnetorotational Instability}.
\newblock {\em \prl}, 97(22):221103, December 2006.

\bibitem{miravet+2022}
Miquel {Miravet-Ten{\'e}s}, Pablo {Cerd{\'a}-Dur{\'a}n}, Martin
  {Obergaulinger}, and Jos{\'e}~A. {Font}.
\newblock {Assessment of a new sub-grid model for magnetohydrodynamical
  turbulence. I. Magnetorotational instability}.
\newblock {\em MNRAS}, 517(3):3505--3524, December 2022.

\bibitem{zhilkin+2021}
A.~G. {Zhilkin} and D.~V. {Bisikalo}.
\newblock {Self-Similar Perturbation of an Accretion Disc around Merging Black
  Holes}.
\newblock {\em Astronomy Reports}, 65(11):1102--1121, November 2021.

\bibitem{lynch+2021}
Elliot~M. {Lynch} and Gordon~I. {Ogilvie}.
\newblock {Dynamical structure of highly eccentric discs with applications to
  tidal disruption events}.
\newblock {\em MNRAS}, 500(3):4110--4125, January 2021.

\bibitem{Mueller+2012}
T.~W.~A. {M{\"u}ller}, W.~{Kley}, and F.~{Meru}.
\newblock {Treating gravity in thin-disk simulations}.
\newblock {\em A\&A}, 541:A123, May 2012.

\bibitem{favre1969}
A~Favre.
\newblock Statistical equations of turbulent gases.
\newblock {\em Problems of hydrodynamics and continuum mechanics}, pages
  231--266, 1969.

\bibitem{papaloizou+1977}
J.~{Papaloizou} and J.~E. {Pringle}.
\newblock {Tidal torques on accretion discs in close binary systems.}
\newblock {\em MNRAS}, 181:441--454, November 1977.

\bibitem{urpin1984}
V.~A. {Urpin}.
\newblock {Hydrodynamic flows in accretion disks.}
\newblock {\em Soviet~Ast.}, 28:50--53, February 1984.

\bibitem{kley+1992}
W.~{Kley} and D.~N.~C. {Lin}.
\newblock {Two-dimensional Viscous Accretion Disk Models. I. On Meridional
  Circulations in Radiative Regions}.
\newblock {\em \apj}, 397:600, October 1992.

\bibitem{regev+2002}
O.~{Regev} and L.~{Gitelman}.
\newblock {Asymptotic models of meridional flows in thin viscous accretion
  disks}.
\newblock {\em A\&A}, 396:623--628, December 2002.

\bibitem{mishra+2022}
R.~{Mishra}, M.~{{\v{C}}emelji{\'c}}, and W.~{Klu{\'z}niak}.
\newblock {Accretion disc backflow in resistive MHD simulations}.
\newblock {\em arXiv e-prints}, page arXiv:2209.06526, September 2022.

\bibitem{hure+2009}
J.~M. {Hur{\'e}} and A.~{Pierens}.
\newblock {A local prescription for the softening length in self-gravitating
  gaseous discs}.
\newblock {\em A\&A}, 507(1):573--579, November 2009.

\end{thebibliography}

%
%%
%%%
%%
%

\end{document}